\documentclass{sf2a-conf2014}
\usepackage{graphicx}
\usepackage{hyperref}
\usepackage[]{natbib}  
\usepackage{epstopdf}
\usepackage{color}

\def\BibTeX{{\rm B\kern-.05em{\sc i\kern-.025em b}\kern-.08em
    T\kern-.1667em\lower.7ex\hbox{E}\kern-.125emX}}
\bibpunct{(}{)}{;}{a}{}{,}  

\begin{document}

\TitreGlobal{SF2A 2014}


\title{2004 -- 2014: Ten years of radiative transfer with STOKES}

\runningtitle{The STOKES code -- a review after ten years}

\author{F.~Marin}\address{Astronomical Institute of the Academy of Sciences, Bo{\v c}ni
			  II 1401, 14131 Prague, Czech Republic}

\author{R.~W.~Goosmann}\address{Observatoire Astronomique de Strasbourg, Universit\'e de Strasbourg, 
				CNRS, UMR 7550, 11 rue de l'Universit\'e, 67000 Strasbourg, France}

\setcounter{page}{237}


\maketitle


\begin{abstract}
Since it became publicly available in 2004, the radiative transfer code {\sc stokes} has been used
to model the spectroscopic, polarimetric, timing and imaging signatures for different astrophysical scenarios. 
Ten years later, at the release of a new version of the Monte Carlo code, we make a census
of the different scientific cases explored with {\sc stokes} and review the main results obtained so far.   
\end{abstract}

\begin{keywords}
Polarization, Radiative transfer, Scattering
\end{keywords}


\section{Introduction}

{\sc stokes} is a Monte Carlo radiative transfer code initially developed by Ren\'e W. Goosmann for his master thesis 
that was supervised by C. M. Gaskell. Since 2010, Fr\'ed\'eric Marin contributes significantly to the development of 
the code and it further benefits from feedback and small extensions implemented by a growing number of users.
Originally created to reproduce  ultraviolet (UV) and optical polarization measurements in radio-quiet active galactic 
nuclei (AGN), the code has evolved toward more versatile versions. The latest public
version of the code, {\sc stokes}~1.2, can be downloaded from the project web page 
\textcolor{blue}{http://www.stokes-program.info/}.

{\sc stokes} is written in $\rm C/C^{++}$ and simulates the radiative transfer with polarization 
for multiple scattering, absorption and/or reemission processes in up to 10$^4$ reprocessing media. The geometry of the 
emitting source and of the scattering structures can be selected from a list of morphologies (toroidal, spherical, 
hourglass-shaped, spherical or more complex, segmented structures) and each region can be characterized by its density, 
temperature and a three-dimensional velocity field. For line emission, Lorentzian flux profiles with user-defined full 
width at half maximum and intensity are assumed. Thomson, Compton and Mie scattering algorithms, using Mueller matrices 
and a Stokes vectorial representation, govern the polarization state of the photon from its emission until its 
eventual escape from the model region. The resulting continuum and line spectra can be evaluated at different inclinations 
and azimuthal viewing angles since the code is fully three-dimensional. The reader is referred to the user manual of 
{\sc stokes} for further details of the code. Several examples of model calculations can also be found on 
the web page.

In this conference note, we give a concise overview of major results achieved with {\sc stokes} for different astrophysical 
cases over the last ten years. Note that some of the following results, especially those related to the X-ray domain, 
are based on more advanced versions than {\sc stokes 1.2} that are not yet public.

\section{Applications and results}

\subsection{Isolated structures}

The first public version 1.0 of {\sc stokes} was described in \citet{Goosmann2007a} and applied to model the effects of different 
AGN reprocessing morphologies on the UV/optical polarized flux. Dusty tori, polar cones and electron disks were studied for 
a range of geometrical shapes, opening angles and optical depths. \citet{Goosmann2007a} found that the shape of the torus funnel 
significantly influences the scattering and polarization efficiency, with compact tori (with a steep inner surface) scattering 
more light along equatorial inclinations than extended tori of the same opening angle. Confined to produce 
optical\footnote{\citet{Goosmann2009} presents the X-ray counterpart of the optical model and illustrates the performance 
of an X-ray polarimeter.} polarization degrees $P$ lower than 20\%, circumnuclear structures are less efficient polarizers than 
polar outflows seen at edge-on (type-2) orientations. While electron-scattering produces wavelength-independent polarization, 
dusty outflows, representative of the so-called narrow line regions (NLR) in AGN, show reddened spectra for polar inclinations 
(type-1) and bluer spectra at type-2 viewing angles.

The case of the NLR has been explored in more detail by \citet{Goosmann2007b,Goosmann2007c}. The authors modeled the polarization 
induced by Mie reprocessing inside a polar, hourglass-shaped structure filled with two different types of dust. The dust 
composition and grain size distribution was based on (1) extinction curves observed in our Galaxy \citep{Mathis1977} and 
(2) a dust mixture inferred from extinction properties of AGN \citep{Gaskell2004}. When the NLR is filled with dust of the 
AGN type, weaker polarization spectra and a net decrease of polarization around 4000~\AA~occur at shorter wavelengths.
This is due to a flatter grain size distribution causing a lower scattering efficiency. As a result, AGN dust scatters less 
flux towards type-2 viewing angles.

\subsection{AGN modeling and the polarization dichotomy}

\subsubsection{The unified model}

Once the polarimetric signatures of individual scattering regions explored, the effects of radiative coupling between the 
inner and outer parts of AGN were included. To do so, three to four reprocessing regions were placed around a central, point 
source: (1) an equatorial, radiation-supported, fully ionized disk producing most of the polarized flux with a polarization 
position angle parallel to the symmetry axis of the AGN, (2) an equatorial, obscuring, dusty torus preventing radiation to 
escape along the midplane, (3) collimated outflows, either filled with electrons (ionized winds) or dust (NLR). \citet{Goosmann2007d}, 
\citet{Marin2011}, \citet{Marin2012c} and \citet{Marin2012d} explored the resulting polarization and stated which constrains 
can be derived from the observed continuum polarization.

It was found that a flat, equatorial reprocessing region with an optical depth of 1 -- 3 is required to generate the observed 
parallel polarization in type-1 AGN, suggesting optically thick accretion flows at the outer edge of type-1 AGN accretion disks. 
Additionally, a wide torus half-opening angle ($\sim$~60$^\circ$) enhances the production of parallel polarization, whereas narrow 
tori and/or a higher optical depth in the polar outflows produce perpendicular polarization when seen at a type-1 viewing angle. At 
type-2 inclinations, all cases modeled produce strong ($\gg$~20\%) perpendicular polarization. This is not in agreement with previous
spectropolarimetric observations of Seyfert-2 galaxies, which report lower ($<$~10\%) polarization percentages. A wide torus/wind 
half-opening angle helps to mitigate the discrepancy with respect to the observations but the resulting type-2 polarization fraction 
remains high. Adding the NLR to the three-component model lowers the production of parallel polarization at type-1 viewing 
angles and helps to decrease the amount of polarization at type-2 viewing angles. A more recent, very careful comparison between 
the models and the observed polarization has been done in \citet[][see also Sect.~\ref{Compendium}]{Marin2014a}.

\subsubsection{Application to a peculiar case: NGC~1068}

The simplest approach of the unified model described in \citet{Antonucci1993} assumes that the accretion disk, the dusty torus 
and the polar outflows are symmetric with respect to the rotational axis of the disk. In this picture, the alignment of the 
ejection winds with the circumnuclear, equatorial matter is due to a collimation effect, while a symmetric mass transfer between 
the inner parts of the torus funnel and the outer edges of the accretion disk would stabilize the two regions along the midplane. 
However, this picture has been questioned by the mid-infrared interferometric measurements of \citet{Raban2009},
who suggest that the polar winds of NGC~1068 (represented as a bi-conical structure) are inclined by 18$^\circ$ with respect to 
the obscuring torus axis.

To investigate the impact of the non-alignment of the ionized winds on the resulting polarization signature of NGC~1068 and, 
by extension, on the unified model of AGN, a multi-wavelength study was carried out with {\sc stokes}. In the X-ray domain, 
\citet{Goosmann2011a,Goosmann2011b} explored a variety of AGN inclinations and different hydrogen column densities of the reprocessing 
material. Under specific conditions, the misalignment of the polar winds with respect to the torus axis can be determined from a 
rotation of the polarization position angle between the soft and the hard X-ray band. In addition to this, soft X-ray polarimetry can 
probe the true orientation of the ionization cones. A similar study in the UV/optical waveband by \citet{Marin2012a} successfully 
reproduces the observed type-1/type-2 polarization dichotomy (i.e. parallel polarization for type-1 AGN and perpendicular 
polarization at type-2 viewing angles) and shows that the polarization is dominated by scattering in the polar outflows and 
therefore traces the wind's tilting angle with respect to the torus axis.

\subsubsection{Observations versus modeling}
\label{Compendium}

Our modeling of NGC~1068 highlights the impact of the system's orientation on the net polarization, in particular for an asymmetric 
geometry. For a given model, changing the orientation of the observer can lead to significantly different results, especially when 
the line of sight is close to the horizon of the circumnuclear torus. It is therefore important to examine a given model at all 
inclinations and to compare the spectropolarimetric results with the observed data. However, this approach was hampered by the lack 
of a data base combining inclination and polarization information until the recent AGN compendium was gathered by \citet{Marin2014a}.

The compendium agrees with past empirical results, i.e. type-1 AGN show low polarization degrees ($P <$ 1\%) predominantly associated
with a parallel polarization position angle while type-2 objects show stronger polarization percentages ($P >$ 7\%)\footnote{Note that 
most, if not all, type-2 AGN polarization measurements from the literature are dominated by relatively large, unpolarized starlight 
fluxes that are insufficiently corrected for (see discussion in Sect.~2.3 of \citealt{Marin2014a}). The  revisited polarization of type-2 
AGN included in the compendium often have lower limits.} with perpendicular polarization angles. The transition between type-1 and 
type-2 inclination occurs between 45$^\circ$ and 60$^\circ$, a range likely including AGN classified as borderline objects, where 
the observer's line of sight crosses the horizon of the equatorial dusty medium. Thanks to this new catalog, the relevance of new 
AGN models can be investigated more properly.

\subsection{The polarization of broad emission lines and off-axis irradiation}

The profile and polarization of broad emission lines can put tight constraints on the geometry of the broad line region. 
The relative offset in blueshift between high and low ionization emission lines could be explained by multiple scattering 
of the line photons in the accretion flow \citep{Gaskell2013}. The technique of {\em polarization reverberation mapping} 
was introduced by \citet{Gaskell2012} as a new way to explore the inner structure of AGN. The cross-correlation between 
the two light curves of the spectral flux and the polarized flux was established and modeled with {\sc stokes} to constrain 
the distance between the continuum source and the reprocessing mirrors. For this purpose, the code was enabled to take 
into account timing properties \citep{Goosmann2008}.

Lately, we started to explore the consequences of a new paradigm for the accretion disk emission. Assuming that a significant 
fraction of the disk's optical/UV luminosity is emitted by temporary off-axis sources (for instance, hot clumps), the 
characteristic polarization profiles across broad emission lines can be explained in a straightforward manner and do not 
require any contribution from the equatorial  scattering disk nor from rotating winds \citep{Goosmann2013}. Observed 
polarization variability on time-scales of the BLR light crossing time would strongly support this new interpretation.

\begin{figure*}
\centering
   \includegraphics[width=14cm]{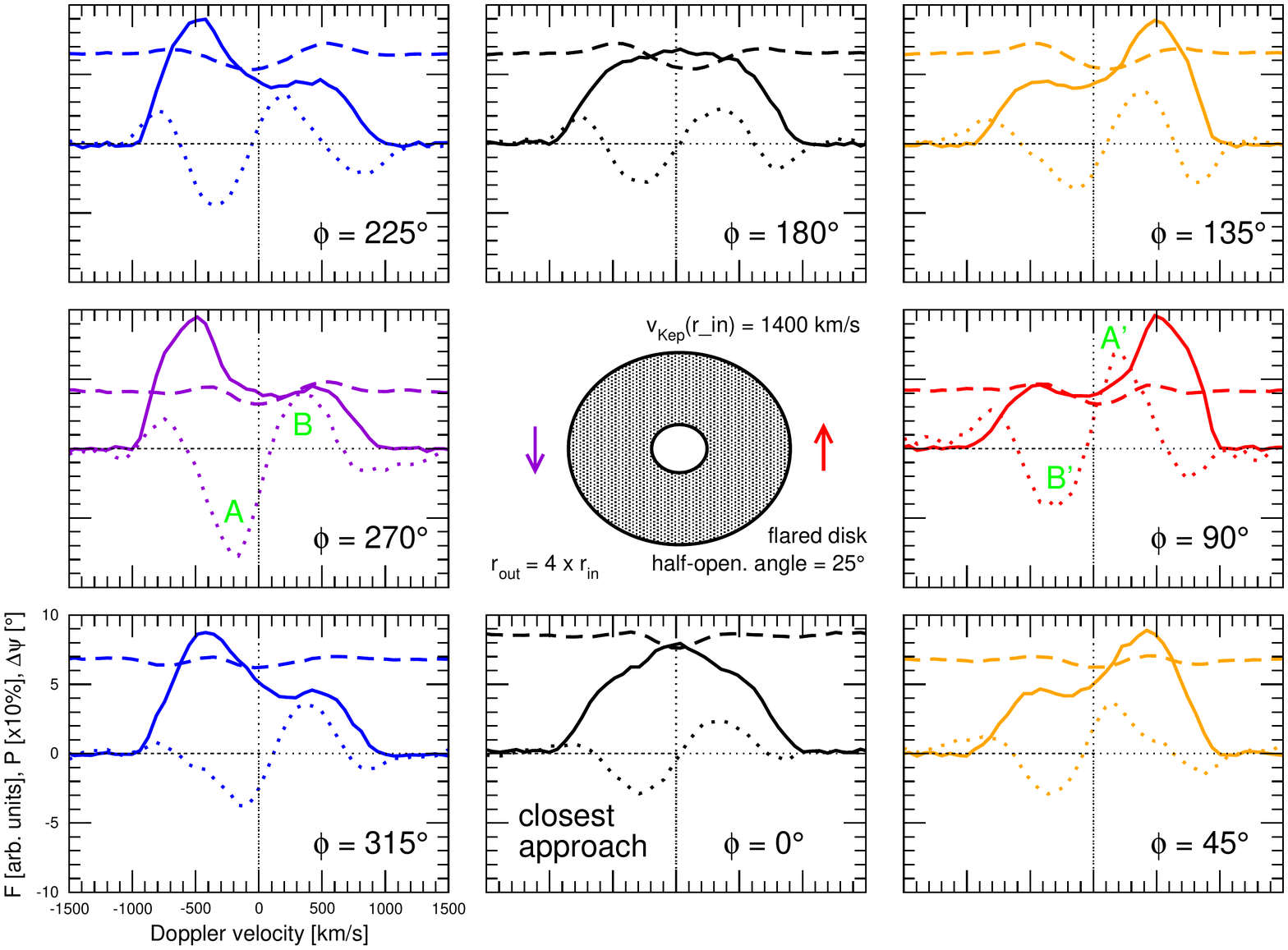}
   \caption{Modeling the spectropolarimetric appearance of a broad emission line as a
	    function of the azimuthal phase $\phi$ with an off-axis continuum source. The 
	    eight panels show the profile of the line flux, $F$ (solid line), the polarization 
	    percentage (dashes), and the variation of the polarization position angle (dots) 
	    in Doppler velocity space. Figure taken from \citet{Goosmann2013}.}
  \label{Fig:OffAxis}
\end{figure*}

\subsection{Constraining the morphology of disk-born outflows}

Despite a wide panel of observational emission and absorption line features to investigate, the morphology of the inner AGN 
regions remains debated. Theories interpreting the reprocessing regions of AGN as dynamical structures propose an alternative 
solution to mostly static media simulated so far \citep{Elvis2000}. In this picture, radiation-driven outflows, potentially 
responsible for a large fraction of AGN feedback, are launched from the accretion  disk over a small range of radii and then 
bent outward and driven into a radial direction by radiation pressure. Along the equatorial plane, shielded from the full 
continuum by the highly ionized matter, dust may survive long enough to create a failed dusty wind.

This scenario is alternative to the torus-based unified model and was tested for its polarization properties by 
\citet{Marin2013b,Marin2013c,Marin2013e}. A model solely composed of ionized winds is unable to reproduce the expected polarization dichotomy 
and underestimates the observed optical polarization percentage of both type-1 and type-2 AGN. A dust-filled outflow produces very low, 
wavelength-dependent polarization degrees, associated with a photon polarization angle perpendicular to the projected symmetry axis of 
the model; the polarization percentages are ten times lower than what can be produced by a toroidal model, with a maximum polarization 
degree found at intermediate viewing angles (i.e. when the observer’s line-of-sight crosses the wind). To agree with the observations,
a two-phase outflow is necessary. It can generate both the observed polarization dichotomy and acceptable levels of polarization degree 
if the wind has a bending angle of 45$^\circ$ (thus lower than what was predicted by the initial, phenomenological model). The conical 
shells need to have a half-opening angle comprised between 3$^\circ$ -- 10$^\circ$ and the absorbing dust column at the wind base 
should correspond to an optical depth (integrated over 2000 -- 8000~\AA) of 1 -- 
4.

\subsection{Probing the physical origin of the asymmetric 6.4~keV iron line}

More recent versions of {\sc STOKES} include radiative processes in the X-ray range. We extended our polarization studies to the 
1 -- 100~keV band and showed that there is important work to be done for X-ray polarimetry. This new window would be an independent 
and complementary tool to spectral and timing analyzes. We focused on a strongly debated topic of X-ray astrophysics, i.e. the 
relativistic reflection versus complex absorption scenarios that are proposed to explain the iron line broadening in a number of 
Seyfert~1 AGN. We managed to derive strong observational predictions for a future spectropolarimetric X-ray mission \citep{Soffitta2013}.

In \citet{Marin2012b}, we modeled the polarization signature of MCG-6-30-15 resulting from a partial covering scenario where a clumpy gas 
distribution is thought to obscure the equatorial plane. We compared the results to a reflection model based on the lamppost geometry and 
found that the shape of the polarization degree and position angle as a function of photon energy are distinctly different between the 
reflection and the absorption cases: disk scattering and general relativistic effects produce significantly stronger polarization in the
soft energy band than absorption. The spectrum of the polarization angle adds additional constraints: its has a constant value in the 
absorption case while smooth rotations of the polarization angle with photon energy are detected in the relativistic reflection scenario. 
We modeled the polarization signature of NGC~1365, a ``changing look'' AGN where variations of cold absorbers on the line of sight cause 
extreme and short emission variability. We showed in \citet{Marin2013d} that a large, soft X-ray observatory or a medium-sized mission 
equipped with a hard (6 -- 35~keV) polarimeter could break the degeneracy between the two scenarios. We summarized and extended our results 
in \citet{Marin2013a}.

\subsection{X-ray mapping of the Galactic Center}

While dedicated to the modeling of AGN, {\sc stokes} can be adapted to a variety of other sources, such as the Galactic Center, 
hosting the closest-to-Earth supermassive black hole. Around the potential well of Sgr~A$^*$ is a concentration of active star formation 
sites and gigantic, reprocessing molecular clouds that make the center of the Milky Way an excellent site for X-ray polarization 
studies. The reprocessing scenario is strongly supported by past X-ray observations of the Eastern massive molecular cloud Sgr~B2 
that revealed a very steep spectrum with a strong emission line at 6.4~keV related to iron fluorescence. This suggests suggesting 
that part of the diffuse emission of Sgr~B2 is due to reprocessing.

\begin{figure*}
\centering
   \includegraphics[trim = 0mm 3mm 0mm 5mm, clip, width=11cm]{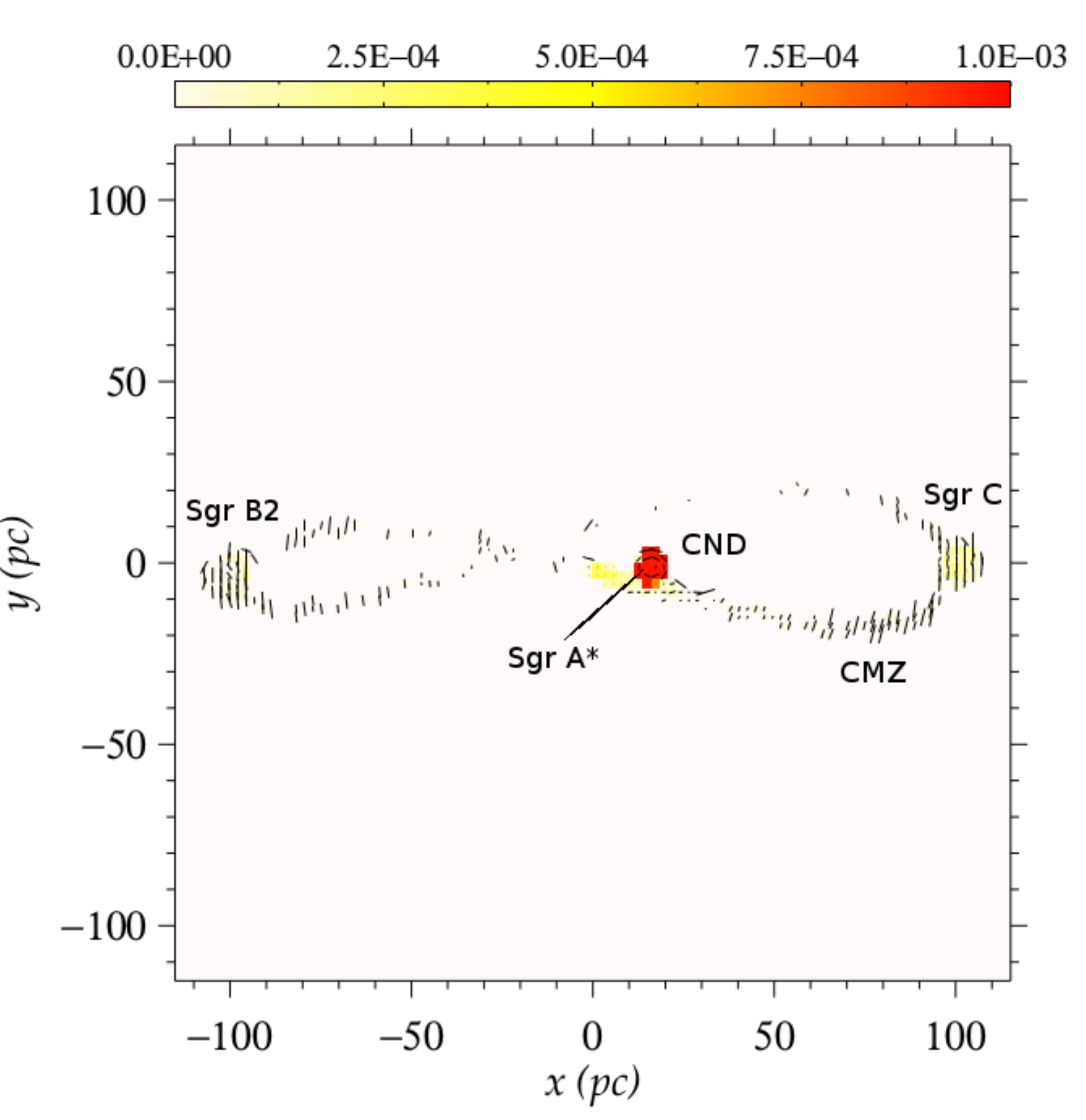}
   \caption{Integrated 8 -- 35~keV model image of the polarized flux, $PF/F_{\rm *}$,
	    for the 2$^\circ~\times~$2$^\circ$ region around the Galactic Center. 
	    $PF/F_{\rm *}$ is color-coded, with the color scale shown on top of the image 
	    (in arbitrary units). The polarization degree and position angle are represented 
	    by black bars drawn in the center of each spatial bin. A vertical bar indicates 
	    a polarization angle of $\psi$~=~90$^\circ$ and a horizontal bar stands for an 
	    angle of $\psi$~=~0$^\circ$. The length of the bar is proportional to $P$. Figure taken from \citet{Marin2014b}}
  \label{Fig:GC}
\end{figure*}

The line-of-sight towards Sgr~A$^*$ being opaque to UV/optical emission, {\sc stokes} was applied in the X-ray range to test if 
a significant X-ray polarization signal could be detected from the Galactic Center. \citet{Marin2014b,Marin2014c} found that a model where 
Sgr~A$^*$ is radiatively coupled to a fragmented circumnuclear disc (CND), an elliptical twisted ring representative of the central 
molecular zone (CMZ), and the two main, bright molecular clouds Sgr~B2 and Sgr~C, produces a variety of polarization signatures. 
Polarization mapping integrated over 8 -- 35~keV reveals that Sgr~B2 and Sgr~C, situated at the two horizontal extensions of the CMZ, 
present the highest polarization degrees, both associated with a polarization position angle normal to the scattering plane. 
The CND shows a lower, barely detectable polarization degree and the CMZ polarization is spatially variable. Independently of their 
spatial location, the two reflection nebulae are found to always produce high polarization degrees ($\gg$~10\%). 
Finally, it has been shown that a 500~ks observation with a broadband imaging polarimeter on-board of a mid-sized mission could 
constrain the location and the morphology of the scattering material with respect to the emitting source.

\section{Concluding remarks}

{\sc stokes} is a versatile Monte Carlo code for modeling polarization produced by absorption, reemission and scattering in many 
astrophysical situations. Its capability to produce UV/optical polarization signatures allows direct comparison with contemporary 
measurements, but is also a prime tool to evaluate the putative X-ray polarization from different environments. Follow-up investigations 
in the infrared band are currently under examination. If you have questions about {\sc stokes}, contact us by email at 
\textcolor{blue}{admin@stokes-program.info}.

\begin{acknowledgements}
The authors thank C.~M.~Gaskell and F.~Tamborra for their involvement in the {\sc stokes} code; the former for being behind the firsts 
lines of the code, the latter for his implication and help in the infrared and X-ray domain. We further acknowledge our colleagues, who
keep using {\sc stokes} and thank them for their feedback and exciting applications.
\end{acknowledgements}

\bibliographystyle{aa}  
\bibliography{marin} 

\end{document}